\documentclass[conference]{IEEEtran}
\IEEEoverridecommandlockouts
\usepackage{cite}
\usepackage{amsmath,amssymb,amsfonts}
\usepackage{algorithmic}
\usepackage{graphicx}
\usepackage{textcomp}
\usepackage{xcolor}

\usepackage{newfloat}
\usepackage{listings}
\usepackage{array}
\usepackage{longtable}
\usepackage{amsmath,amsthm,amssymb,amsfonts}

\usepackage{booktabs}
\usepackage{multirow}
\usepackage{siunitx}
\usepackage{pifont}
\usepackage{caption}
\usepackage{subcaption}
\usepackage[colorlinks=true, urlcolor=cyan]{hyperref}
\def\BibTeX{{\rm B\kern-.05em{\sc i\kern-.025em b}\kern-.08em
    T\kern-.1667em\lower.7ex\hbox{E}\kern-.125emX}}
\begin{document}

\title{ECG-Chat: A Large ECG-Language Model for Cardiac Disease Diagnosis}

\author{\IEEEauthorblockN{Yubao Zhao\IEEEauthorrefmark{3}\IEEEauthorrefmark{1},
Jiaju Kang\IEEEauthorrefmark{4}\IEEEauthorrefmark{1}\IEEEauthorrefmark{2},
Tian Zhang\IEEEauthorrefmark{5}, 
Puyu Han\IEEEauthorrefmark{6}, and
Tong Chen\IEEEauthorrefmark{7}}
\IEEEauthorblockA{\IEEEauthorrefmark{3}China University of Geosciences \IEEEauthorrefmark{4}Beijing Normal University \IEEEauthorrefmark{5}ESIGELEC }

\IEEEauthorblockA{\IEEEauthorrefmark{6}Southern University of Science and Technology \IEEEauthorrefmark{7}University of Liverpool}

\IEEEauthorblockA{\tt\small yubaozhao01@gmail.com, kjj\_python@163.com}
\thanks{\IEEEauthorrefmark{1}Equal contribution. \IEEEauthorrefmark{2}Corresponding author.}}

\maketitle

\begin{abstract}
   The success of Multimodal Large Language Models (MLLMs) in the medical auxiliary field shows great potential, allowing patients to engage in conversations using physiological signal data. However, general MLLMs perform poorly in cardiac disease diagnosis, particularly in the integration of ECG data analysis and medical report generation, mainly due to the complexity of ECG data analysis and the gap between text and ECG signal modalities. To address these issues, we propose ECG-Chat, a multitask MLLMs focused on ECG medical report generation, providing multimodal conversational capabilities based on cardiology knowledge. We propose a contrastive learning approach that integrates ECG waveform data with text reports, aligning ECG features with reports in a fine-grained manner. This method also results in an ECG encoder that excels in zero-shot report retrieval tasks. Additionally, expanding existing datasets, we constructed a 19k ECG diagnosis dataset and a 25k multi-turn dialogue dataset for training and fine-tuning ECG-Chat, which provides professional diagnostic and conversational capabilities. Furthermore, ECG-Chat can generate comprehensive ECG analysis reports through an automated LaTeX generation pipeline. We established a benchmark for the ECG report generation task and tested our model on multiple baselines. ECG-Chat achieved the best performance in classification, retrieval, and medical report generation tasks. Our code is available at \url{https://github.com/YubaoZhao/ECG-Chat}.
\end{abstract}

\begin{IEEEkeywords}
ECG, Large Language Model, Multimodal
\end{IEEEkeywords}

\section{Introduction}


With the rapid development of multimodal large language models (MLLMs), these advanced systems show great promise in assisting patients by integrating diverse modalities such as text, images, and medical data to enhance healthcare delivery and decision-making. 
As a non-invasive physiological indicator detection method, electrocardiography (ECG) is a crucial tool for detecting early heart problems in patients. Some existing algorithms treat ECG data as a temporal physiological signal, focusing mainly on classification tasks \cite{markov2023,xu2024, davies2024, hoang2024}. These models cannot help patients and can only serve as auxiliary tools for doctors. Also, Current ECG-language models 
 \cite{liu2024b, wan2024electrocardiogram} have not bridged the gap between ECG and text report modalities because most open ECG datasets lack comprehensive text descriptions. The structured, terminological, and highly repetitive phrase combinations poses challenge for the current Vision-Language models to migrate to the ECG field. Consequently, no model currently addresses both ECG report generation and question answering effectively. Additionally, large language models suffer from significant hallucinations in cardiology, making their accuracy unreliable.

In this paper, to address these challenges, we introduce ECG-Chat, a MLLM capable of generating long text reports for ECG. ECG-Chat uses the ECG encoder trained on the framework of contrastive learning. The original report is enhanced with waveform data, which shows excellent performance in retrieval tasks. After that, we used GPT-4o to build a dataset, ECG-Instruct, for instruction tuning ECG-LLM, which includes two forms: diagnosis and dialogue. With the ECG encoder, dataset and the LLM Vicuna-13B \cite{vicuna}, we constructed ECG-Chat that supports multiple functions such as report generation and ECG question and answer. At the same time, in order to solve the model's hallucinations in report generation and medical knowledge, we built a prompt template for specific ECG diagnoses and a local knowledge base for retrieval enhancement generation (RAG). In addition, we built a pipeline that integrates patient information to generate detailed ECG reports. We tested the performance of the model on multiple tasks such as ECG-Report retrieval, ECG classification and ECG report generation. And established a benchmark for LLMs in ECG report generation.

In summary, this paper makes the following contributions:

\begin{itemize}
    \item \textbf{Waveform Data Enhancement}: Introduced a contrastive learning method aligning ECG waveform data with text reports, achieving state-of-the-art ECG encoder performance. 
   \item \textbf{ECG-Instruct}: Developed a novel pipeline using GPT-4 to generate a 19K-diagnosis and 25K-dialogue ECG instruction tuning dataset. 
   \item \textbf{ECG-Chat}: Fine-tuned Vicuna-13B with LoRA to prevent catastrophic forgetting, aligning ECG signals with LLM text embeddings. Proposed a Diagnosis-Driven Prompt (DDP) for accurate ECG report generation, incorporating automated LaTeX-based medical vocabulary explanations for comprehensive and accessible reports. 
\end{itemize}
\begin{figure*}[t]
	\centering
	\includegraphics[width=2\columnwidth]{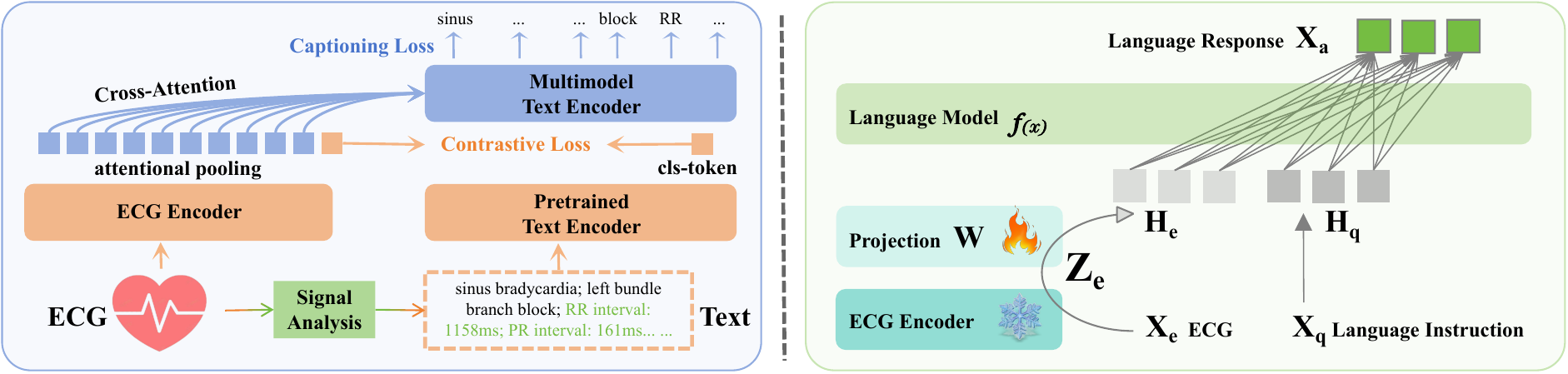} 
	\caption{The diagram depicts training process for ECG encoder (left part) and ECG-Chat (right part).}
	\label{fig2}
        \vspace{-1mm}
\end{figure*}

\section{Related Work}
\subsection{Cross-modal Medical Text Generation}
Cross-modal medical text generation refers to the process of leveraging multiple data modalities, such as images, text, and signals, to generate accurate and medically informed diagnostic or explanatory text, thereby supporting medical decision-making and treatment processes. Certain studies have successfully integrated multiple tasks into a single model, demonstrating remarkable functionality \cite{zhang2023}. As the field has progressed, unstructured data has been incorporated. \cite{wan2024, gupta2024, yu2024, chen2024} Notably, image-text multimodal technology has shown exceptional performance in this domain, significantly enhancing diagnostic accuracy and medical efficiency. \cite{zhang2022} However, research on signal-text multimodality, particularly in applications involving electrocardiograms, remains underexplored and holds substantial potential for future development.
\subsection{Intelligent Interpretation of ECG}
The intelligent interpretation of ECG encompasses a range of tasks, including ECG retrieval, classification, and text generation, aimed at enhancing the accuracy and efficiency of cardiac diagnosis. In ECG retrieval, advanced algorithms are employed to search and identify relevant ECG patterns from large databases, enabling efficient comparisons and aiding in clinical decision-making\cite{qiu2023}. ECG classification has seen significant progress, with machine learning models being developed to automatically categorize ECG signals into various cardiac conditions, demonstrating high accuracy in identifying abnormalities\cite{yu2024b,liu2024b, markov2023,xu2024, davies2024, hoang2024, plagwitz2024, huang2024}. Despite these advancements, the generation of descriptive and diagnostic text from ECG data remains a relatively nascent area of research \cite{xu2024,jiang2024, wan2024electrocardiogram}. Current studies focus on leveraging multimodal approaches to generate text that accurately reflects the underlying ECG signals, providing clinicians with clear and actionable insights.  However, challenges such as ensuring the clinical relevance and interpretability of generated text, as well as addressing the nuances of complex cardiac conditions, remain areas requiring further exploration.


\section{Method}

\begin{figure*}[t]
	\centering
	\includegraphics[width=2\columnwidth]{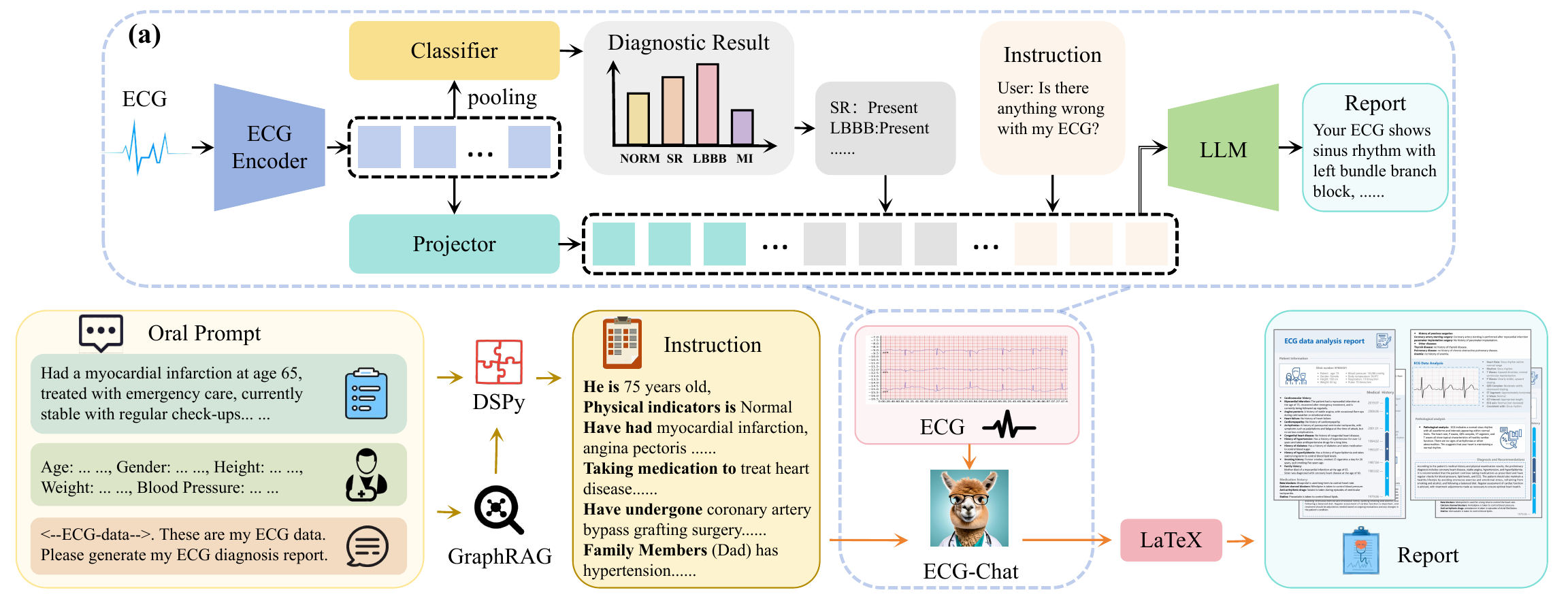} 
	\caption{ECG-Chat improves long-form medical report generation by integrating automated prompt tuning and GraphRAG into its system. This approach extends to other chat functions as well. As illustrated in (a), a key component is a classifier  (shown as gray blocks), built as a linear layer, trained to predict the likelihood of various diagnoses. }
	\label{fig_ddp}
    \vspace{-1mm}
\end{figure*}

\subsection{Architecture and General Pipeline}
ECG-Chat contains an end-to-end ECG diagnostic report generation pipeline. The ECG signals firstly pass through an ECG encoder, transforming the time series into feature representations. To enable the features fine-grained aligned with reports, we followed the contrastive learning architecture of the CoCa \cite{coca}, and enhanced the text report with ECG waveform data. After that, similar to LLaVa-v1.5 \cite{improved_llava}, we use a two-layer MLP adapter to align the feature space of the ECG encoder with the LLM through pretraining and fine-tuning. At the same time, we built a Dignosis-Driven Prompt based on linear classification for accurate ECG report generation, and built GraphRAG to address hallucinations in medical knowledge. Finally, a large language model of ECG diagnosis with the ability of report generation and multi-round conversations is constructed. 
\subsection{Aligning ECG Features with Text Reports}
In visual multimodal large language models, visual encoders are often obtained by training on large-scale image-text datasets by dual-encoder contrastiva learning\cite{clip} or encoder-decoder captioning \cite{simvlm}. CoCa \cite{coca} is a model that combines the two approches. It maps image to the same space as text representations by optimizing the contrastive loss through a dual-encoder. At the same time, the multimodal text decoder is used to optimize the captioning loss and improve the cross-modal generation ability of the model. Inspired by CoCa's success in computer vision and ECG classification \cite{yu2024b}, we also use this model to extract features from ECG signals. Our architecture similarly uses contrastive loss and captioning loss, as shown in Fig. \ref{fig2}.

Optimizing the contrastive loss is the way that ECG features align with the text reports, represented in a batch of training samples as follows:
\begin{align}\label{con_loss_e2t}
       \mathcal{L}_{e2t} &=\sum_{i}^{N}\log{\frac{\exp (x_{i}^{T}y_{i}/\sigma )}{ {\textstyle \sum_{j=1}^{N}\exp (x_{i}^{T}y_{j}/\sigma )} } }\\
\label{con_loss_t2e}
       \mathcal{L}_{t2e} &=\sum_{i}^{N}\log{\frac{\exp (y_{i}^{T}x_{i}/\sigma )}{ {\textstyle \sum_{j=1}^{N}\exp (y_{i}^{T}x_{j}/\sigma )} } }\\
\label{con_loss}
       \mathcal{L}_{con} &=-\frac{1}{N}(\mathcal{L}_{e2t}+\mathcal{L}_{t2e})
\end{align}
where \(\mathcal{L}_{e2t}\) and \(\mathcal{L}_{t2e}\) the contrastive loss of ecg-to-text and text-to-ecg, respectively. \(x_{i}\) and \(y_{i}\) are normalized embeddings of the ECG signal in the \(i\)-th pair and that of the text in \(j\)-th pair. \(N\) is batch size, and \(\sigma\) is the temperature to scale the logits. To speed up the training, we use a pretrained text encoder and fine-tune only its last few layers \cite{li2024}.

Optimizing the captioning loss is to make the ECG features predict the exact tokenized texts of \(y\) in an auto-regressive way. ECG encoder provides the latent ECG features and the text decoder learn to maximize the conditional likelihood of a pair of texts y under the forward autoregressive factorization:
\begin{equation}\label{cap_loss}
       \mathcal{L}_{cap} =-\sum_{t=1}^{T}\log P_{\theta}(y_{t}|y_{<t}, x)
\end{equation}
where \(x\) is the ECG latent feature, \(\theta\) is the parameters of ECG encoder and multimodal text decoder. Then, the loss function of our model can be expressed as:
\begin{equation}\label{coca_loss}
       \mathcal{L} = \lambda_{con}\cdot\mathcal{L}_{con}+
       \lambda_{cap}\cdot\mathcal{L}_{cap}
\end{equation}
where \(\lambda_{con}\) and \(\lambda_{cap}\) are loss weighting hyper parameters.

For the training data, we did not use diagnostic reports directly. Given the scarcity of datasets with text reports, we appended corresponding waveform data to the reports (Fig. \ref{fig2}). This approach artificially increases the distinction between samples, even when reports are identical, helping prevent contrastive loss from failing to converge during small-batch training. Additionally, incorporating waveform data ensures that the ECG encoding latent space captures more waveform information.

\subsection{Multimodal Instruction Turning and Inference}
The modality interface in ECG-Chat resembles LLaVA-v1.5 \cite{improved_llava}. ECG encoding is embedded like text tokens and fed into LLMs. Since ECG and text feature spaces differ, an adapter—a two-layer MLP—is used for conversion. ECG-Chat undergoes two-stage pretraining: feature alignment and instruction fine-tuning, with the ECG encoder frozen in both stages. During feature alignment, only the linear projection layer is trained, while the LLM remains frozen. In the fine-tuning stage, the LLM is trained with LoRA \cite{lora}. Due to the lack of existing datasets, we generated pretraining data and the ECG-Instruct dataset using GPT-4o, which will be detailed in the following chapter.

Large language models struggle with generating precise medical reports. Inspired by the Diagnosis-Driven Prompt (DDP) method in radiology \cite{promptmrg, chen2024diallama}, we adapted DDP for ECG report generation (Fig. \ref{fig_ddp}). A linear layer classifies ECG feature vectors, embedding the most probable label in the prompt (e.g., \textit{"The {label description} is present."}). For multi-label ECG classification, we enhance accuracy by grouping labels into disease, rhythm, and waveform categories. Labels with probabilities above a threshold are included, treating rhythm as single-label. For example, in Fig. \ref{fig_ddp}, outputs like sinus rhythm (SR) and left bundle branch block (LBBB) are embedded as \textit{"Sinus rhythm is present; Left bundle branch block is present"} in the model input.

\subsection{Hallucination Elimination and Visualization Output}

Our approach integrates three essential components to enhance the generation of ECG diagnostic reports. First, the GraphRAG component constructs a comprehensive knowledge graph from seven authoritative cardiology textbooks, enabling effective Retrieval-Augmented Generation (RAG) to mitigate hallucinations. This ensures that the generated text is grounded in established medical knowledge. Second, DSPy is employed for automated prompt tuning, which dynamically integrates patient data with retrieved knowledge, producing accurate and contextually relevant outputs. Finally, the LaTeX-based pipeline automates the creation of structured clinical reports, ensuring clarity and consistency in the final output. Detailed descriptions of these components are provided in Appendices A through C.

\section{Dataset}
\subsection{Datasets for Contrastive Learning}
In the framework of ECG-Chat, the ECG encoder is trained using a large number of ECG-text pairs of data by contrastive learning. The existing open source datasets are relatively few, and the data size is also very small. Current multimodal models often integrate multiple datasets to improve generalization \cite{clip}. Therefore, we also integrate three datasets for ECG encoder pre-training, namely MIMIC-IV-ECG \cite{mimic-ecg}, Champan-Shaoxing-Ningbo (CSN) \cite{csn_data1, csn_data2} and Shandong Provincial Hospital (SPH) \cite{sph_data} datasets. The total size of the training dataset is 805K. For detailed data set description and preprocessing, please see the Appendix D.

For the waveform data enhanced report, we used Python toolbox NeuroKit2 \cite{neurokit} to extract the waveform information of each recording on lead II separately, including RR interval, PR interval, QRS complex duration, QT/QTc interval, and the peaks of P, R and T waves. The waveform data is added directly after the text report, as shown in the Fig. \ref{fig2}.

\begin{figure}[h]
        \centering
	\includegraphics[width=\columnwidth]{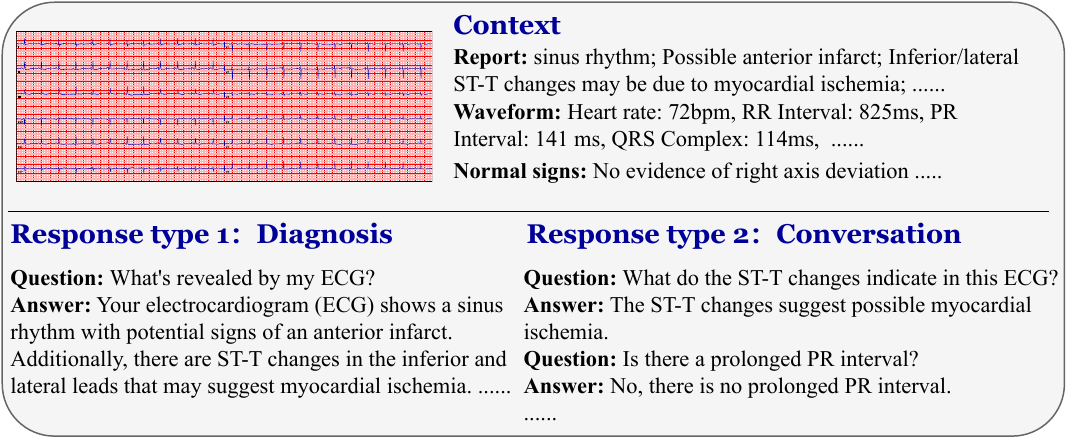} 
	\caption{An example of the ECG-Instruct dataset. The context is used as the prompt of GPT-4 to describe an ECG.}
	\label{fig_ecg_ins}
\end{figure}

\subsection{Datasets for Instrution Tuning}
ECG-Chat is trained in two stages. The training data for both are obtained from MIMIC-IV-ECG \cite{mimic-ecg}. The pretraining dataset includes a 619K records, each record has a random GPT-4 generated user instruction, and the reports in the dataset constitute the LLM's answer. The fine-tuning dataset includes 19K diagnosis data and 25K conversation data. The diagnosis data is similar to the pretraining data, but the answer template is more in line with user interaction. The conversation data set contains 4-15 rounds of dialogue about ECG details. The context and training data when constructing the fine-tuning dataset are shown in Figure \ref{fig_ecg_ins}. The specific construction process is shown in Appendix E.

\subsection{Datasets for Evaluations}
For ECG encoder, we evaluate on ECG-Report retrieval and zero-shot, linear probe classification. For ECG-Chat, we evaluate on ECG report generation task. The datasets used for evaluation are PTB-XL \cite{ptbxl} and CPSC2018 \cite{cpsc2018} datasets. For the ablation study of GraphRAG and DSPy, we use the ECG-ExpertQA dataset, which is constructed by chatGPT-4o. More information can be found in Appendix D.

\section{Experiment}
\subsection{Implementation Details}
In the ECG encoder training phase, we used a 12-layer 1d-ViT as the backbone. ECG-ViT has a patch size of 50, a hidden size of 768, 12 heads, and an MLP size of 3072. For the text encoder, we used the model Med-CPT \cite{medcpt} obtained on the text contrastive learning task and only fine-tuned its last two layers. The embedding dimension of the dual encoder is 512. The text decoder is also a 6-layer transformer decoder. During training, the batch size per GPU is 128. The AdamW optimizer with a learning rate of 1e-4 and a weight decay of 0.1 is used for 20 epochs. The weights of contrastive loss and captioning loss are 1.0 and 2.0 respectively. In addition, we used three ECG data augmentation strategies: baseline wander, cut mix and random masking \cite{torch_ecg}.

In ECG-Chat, the projection layer is a two-layer MLP, and the LLM base is Vicuna-13B \cite{vicuna}. AdamW optimizer with a cosine learning rate scheduler is used to train our model. The pretraining stage lasts for 1 epoch, where only the projection layer is trained. The fine-tuning stage lasts for 3 epochs, where the projection layer is trained and Vicuna-13B is fine-tuned with LoRA \cite{lora}.In order to save memory, ZeRO \cite{zero} was used in the construction of ECG-Chat.

All experiments were run on 8×V100 32GB GPUs.
\subsection{Zero-shot ECG-Report Retrieval}
ECG-report retrieval is a cross-modal task that involves using an ECG or text modality to find the matching modality from a database. We evaluate the ECG encoder's feature alignment ability by encoding ECG and text reports separately, then calculating similarity between their feature vectors and those in the database. The record with the highest similarity is retrieved. For testing, we use ECG records from the PTB-XL test dataset and English-translated free text reports. This dataset helps assess the model's generalization performance due to its different report style. We evaluate using Recall at K (R@K) with K=1, 5, 10.

\begin{table}[h]
 \small
 \centering
 \setlength{\tabcolsep}{4pt}
 \renewcommand{\arraystretch}{1.2}
  \caption{Zero-shot ECG-report retrieval results on PTB-XL (2K test set)}
 \begin{tabular}{cccc|ccccc}
 \toprule
 & \multicolumn{3}{c}{ECG to Report} & \multicolumn{3}{c}{Report to ECG} \\
 \cmidrule(lr){2-4} \cmidrule(lr){5-7}
 Model & R@1 & R@5 & R@10 & R@1 & R@5 & R@10 \\
 \midrule
 All-Grid & 0.21 & 1.06 & 1.91 & 0.43 & 1.06 & 1.91 \\
 MERL & 1.00 & 2.91 & 5.23 & 0.96 & 3.28 & 5.37 \\
 ALBEF & 1.36 & 3.46 & 5.78 & 1.00 & 3.50 & 5.64 \\
 CoCa & 2.14 & 6.65 & 9.60  & 2.37 & 6.10 & 10.2 \\
 \midrule
 CoCa+WDE & \textbf{64.7} & \textbf{84.7} & \textbf{89.4} & \textbf{71.6} & \textbf{89.0} & \textbf{93.0}\\
 \bottomrule
 \end{tabular}

 \label{tab_retrival}
\end{table}


\begin{table*}[t]
 \centering
 \setlength{\tabcolsep}{2pt}
 \renewcommand{\arraystretch}{1.2}
     \caption{Ablation study of DDP in clinical efficacy (CE) and natural language generation (NLG) metrics evaluation}
    \begin{tabular}{lc|ccccccccc|cccc}
        \toprule
         & \multicolumn{1}{c}{ }& \multicolumn{3}{c}{CE-Disease} & \multicolumn{3}{c}{CE-Form} & \multicolumn{3}{c}{CE-Rhythm} & \multicolumn{4}{c}{NLG}\\
        \cmidrule(lr){3-5} \cmidrule(lr){6-8} \cmidrule(lr){9-11} \cmidrule(lr){12-15} 
        & DDP & Pre. & Rec. & F1 & Pre. & Rec. & F1 & Pre. & Rec. & F1 & BLEU-1 & BLEU-4 & ROUGE-L & METEOR\\
        \midrule
        PTB-XL & - &- & - & - & - & - & - & - & - & - & 6.48 & 0.88 &  25.62 & 17.23\\
        ECG-Chat &\ding{56} &3.40 & 4.00 & 1.76 & 1.53 & 4.10 & 0.98 & 6.33 & 5.74 & 13.04 & 15.91 &  2.32 & 23.87 & 29.39\\
        ECG-Chat &\ding{52} &\textbf{33.60} & \textbf{18.91} & \textbf{22.33} & \textbf{25.54} & \textbf{15.11} & \textbf{17.35} & \textbf{54.76} & \textbf{40.02} & \textbf{43.39} & \textbf{32.27}  & \textbf{11.19}  & \textbf{29.93} & \textbf{35.10}\\
        \bottomrule
    \end{tabular} 

    \label{tab_report_ce}

\end{table*}

Table \ref{tab_retrival} shows the retrieval results of ECG-Chat’s ECG encoder on the PTB-XL test set of size 2K, and compares it with models that do not use Waveform Data Enhancement (WDE), including All-Grid \cite{qiu2023b} 
 which using image encoding, ALBEF \cite{ALBEF}, MERL \cite{liu2024b}, and CoCa \cite{coca} models. ALBEF, MERL, and CoCa use the same ECG ViT encoder, pretrained text encoder and training datasets as our model. As can be seen from Table \ref{tab_retrival}, the CoCa model achieves the best results in retrieval. Also, without WDE, the recall of the retrieval is very low. WDE increases the differences between ECG reports and provides more information. The CoCa model with WDE achieves the best results.




\subsection{ECG Reports Generation}
To evaluate the diagnostic capability of ECG-Chat, we conducted an evaluation on the report generation task. Due to the completeness of the PTB-XL dataset labels, we still use its test set. During the evaluation, the user's instructions are unified as \textit{"Could you please help me explain my ECG?"}. For DDP, we classify ECG records according to 3 groups of non-overlapping labels: Disease, Form, and Rhythm. The three groups of labels are all SCP codes used in PTB-XL. Among them, the Disease corresponds to the 40 categories of the Subclass that do not overlap with the Form, and the Form and Rhythm are consistent with the definitions in the PTB-XL dataset. The classifier is a series of binary classifiers trained on the training set of PTB-XL.

\begin{figure}[t]
        \centering
	\includegraphics[width=1\columnwidth]{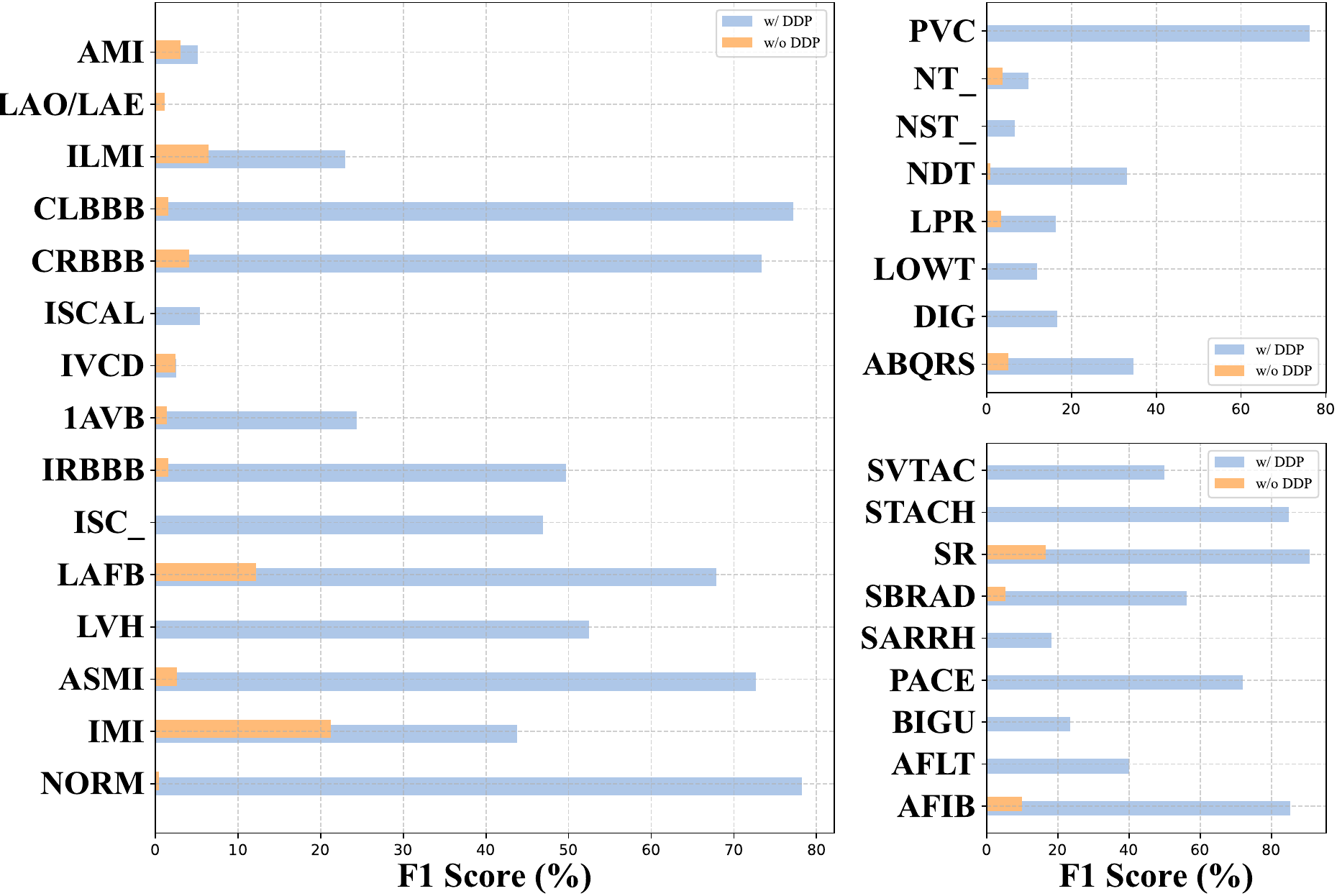}

 \caption{ECG-Chat F1 score statistics for different categories. The left side is the Disease subset, and the right side is the Form and Rhythm subsets. The labels are abbreviated as SCP Codes\cite{scp_ecg}. }
    \label{fig_report_f1}

\end{figure}

For evaluation, both clinical efficacy (CE) and natural language generation (NLG) metrics are used. CE includes Precision, Recall and F1 score, where the labels of text reports are annotated by GPT-4. NLG includes BLEU, METEOR  and ROUGE-L. NLG meteics uses GPT-4o’s answers based on PTB-XL translated reports as reference.

Table \ref{tab_report_ce} shows the NLG evaluation results of ECG-Chat. We compare it with the translated original report in PTB-XL. ECG-Chat with DDP achieves the best results in all indicators. The original report in PTB-XL is shorter, so it is not as good as ECG-Chat in BLUE and METEOR metrics. At the same time, ECG-Chat is also better than the original report in ROUGE-1 and ROUGE-L, which may be because ECG-Chat has the same template as the reference GPT-4o response, and DDP also greatly improves the accuracy of the answer.

Table \ref{tab_report_ce} also shows the results of DDP on the CE metrics. Due to the small number of training data sets, the model without DDP showed serious hallucinations. Therefore, at the current stage, this prompt is indispensable. However, on the three data sets, the recall is relatively low, indicating that the model has some problems in judging negative samples.

Figure \ref{fig_report_f1} shows the classification F1 scores of some labels. Some common labels, such as "Normal (NORM)", "Sinus rhythm (SR)", have high F1 scores. For uncommon labels, the accuracy is very low. Many labels have an F1 score of 0. Moreover, without DDP, the model only prefers a few specific responses. Appendix G shows several reports generated by ECG-Chat. Overall, ECG-Chat is able to generate relatively accurate ECG interpretation reports.


\section{Conclusion and Limitations}

We introduced ECG-Chat, the first large ECG-language model for cross-modal cardiac diagnosis. To address the lack of instruction tuning datasets, we also released ECG-Instruct, a multi-task, multi-modal dataset. ECG-Chat achieves state-of-the-art results in ECG medical text generation by applying WDE and DDP. WDE enhances the ViT encoder, validated through ECG-Text retrieval and classification tasks, while DDP corrects report generation errors with information from classifiers. We utilized GraphRAG and DSPy to refine the model's self-optimization and tackle hallucination issues in medical text generation. ECG-Chat highlights the importance of multi-stage information fusion and the potential of large models in medical scenarios. We hope ECG-Chat and ECG-Instruct will advance ECG representation learning.

Although our model has achieved excellent results on multiple tasks, there are still some limitations. First, due to the lack of diversity in the training dataset, the ECG features cannot be aligned with the LLM model as expected. At the same time, the dataset for instruction tuning is small and does not come from real world, which leads to bias and hallucinations in LLMs. Second, the model can still be improved in diagnosis, especially in waveforms and some rare symptoms. Finally, our model only focuses on ECG, and we also hope to combine it with the most advanced medical large language model \cite{chen2024huatuogpt}, integrating multiple modalities such as X-Ray, EHR, etc., to provide a more comprehensive interpretation of medical records.

\bibliographystyle{IEEEbib}
\bibliography{icme2025references}

\clearpage
\appendices
\section{RAG of Cardiology Knowledge}
Severe hallucination issues are a major constraint in generating medical reports using large models, manifested as semantic deviations caused by generated text that does not align with the actual situation. The root cause is often the model's lack of prior knowledge in certain specialized fields. One approach to supplementing the prior knowledge of large models involves fine-tuning the model by retraining it on new datasets to update its knowledge. However, this method is prone to catastrophic forgetting, incurs unacceptable training costs, and exhibits poor scalability when dealing with dynamic data. In contrast, Retrieval-Augmented Generation (RAG) uses external knowledge bases and retrieval algorithms, enabling LLMs to generate relevant responses based on previously unseen data, effectively overcoming the hallucination problem in LLMs.

ECG-Chat uses Microsoft's GraphRAG \cite{edge2024localglobalgraphrag} component to convert the content from professional books such as "ECG Workout - Exercises in Arrhythmia Interpretation \cite{ecg_workout_8th}", "Manual of Cardiovascular Medicine \cite{manual_cardiovascular_medicine_4th}", "Medical Student Survival Skills ECG \cite{medical_student_survival_skills_ecg}", "Cardiology Subspecialty Consult \cite{cardiology_subspecialty_consult}", "The ECG Made Easy \cite{the_ecg_made_easy_50th}", "The ECG In Practice \cite{the_ecg_in_practice}", and "Arrhythmia Recognition: The Art of Interpretation 
\cite{arrhythmia_recognition}" into a graph index. This index, built through a knowledge graph of nodes and edges, comprehensively covers the knowledge of ECG interpretation and cardiovascular diagnosis. GraphRAG first parses the content of these books into a graph structure and uses community detection algorithms to group related medical topics. When a user submits a query, the system retrieves and summarizes relevant elements from the graph index, generating a comprehensive "global answer," thereby helping to mitigate hallucinations in LLMs during the generation of ECG diagnostic reports. Ablation experiments have validated that this module is effective.
\section{Automated Prompt Tuning}
Diagnosing complex medical conditions and explaining intricate medical terminology often pose challenges in achieving satisfactory results within a single conversation handled by LLMs.  In many cases, prompts are empirically developed by developers through repeated attempts and then fixed.  However, LLMs are sensitive to prompts, leading to fragility in practical applications.

We employ a DSPy component \cite{soylu2024finetuningpromptoptimizationgreat}, trained on the ECG-Instruct dataset, to perform automatic prompt tuning.  DSPy leverages GraphRAG to retrieve relevant knowledge such as symptom causes, medication information, and clinical guidelines, and combines this knowledge with the patient’s vital signs and cardiology diagnostic data to form a comprehensive medical context.  Based on this, the DSPy module automatically generates accurate and detailed medical diagnostic texts through a series of declarative programming steps. The automation and modular design of this workflow enable DSPy to optimize the prompts and weights of the language model, improving the accuracy and efficiency of the output while reducing the need for manual tuning, thereby enhancing the system's scalability and maintainability.

\section{LaTeX-based Automated Medical Report Generation Pipeline}
\begin{figure*}[htb]
    \centering
    \begin{minipage}[b]{0.46\textwidth}
        \centering
        \includegraphics[width=\textwidth]{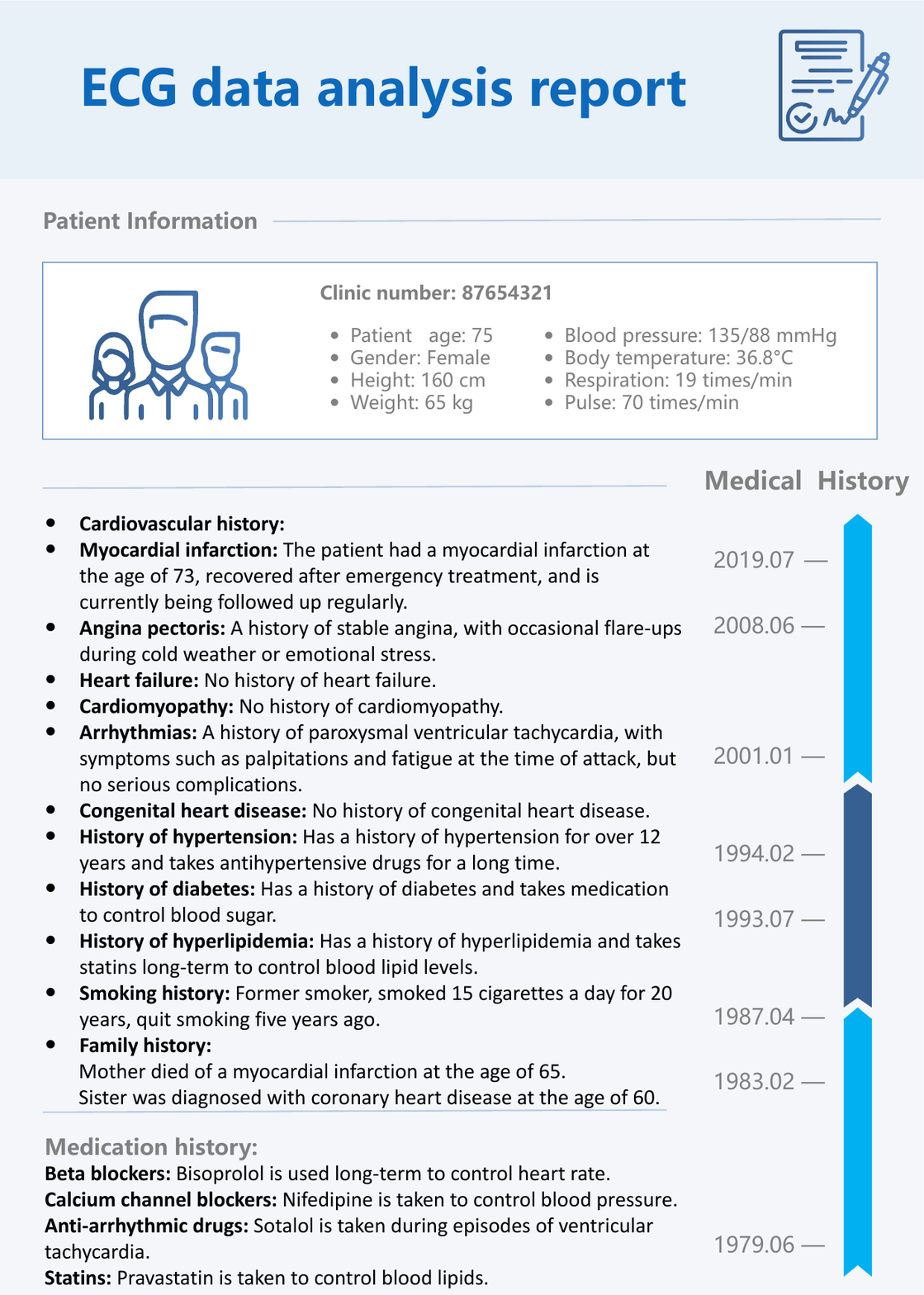}
        \caption{Page 1}
        \label{fig:report_1}
    \end{minipage}
    \hfill
    \begin{minipage}[b]{0.46\textwidth}
        \centering
        \includegraphics[width=\textwidth]{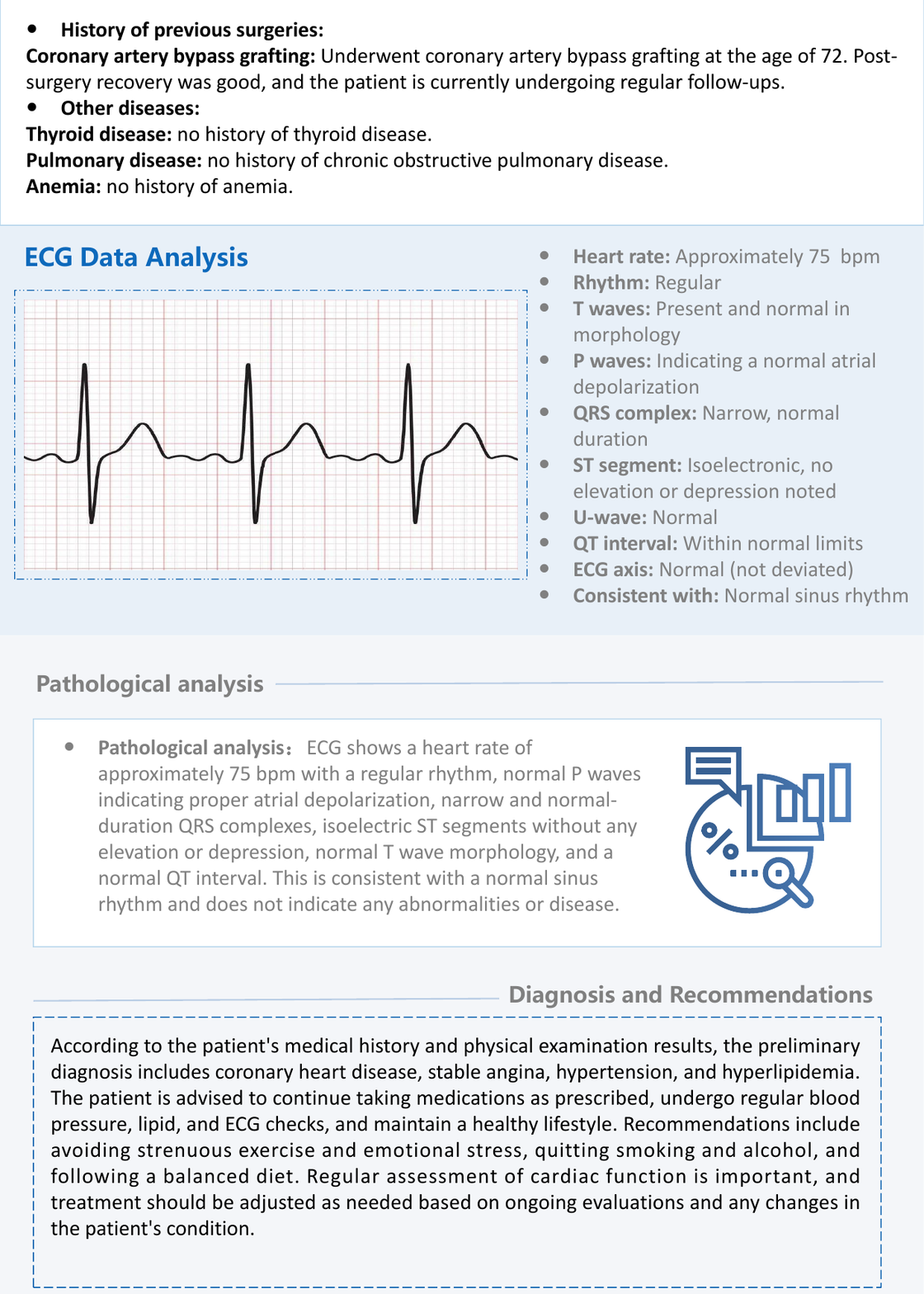}
        \caption{Page 2}
        \label{fig:report_2}
    \end{minipage}
\end{figure*}
Clinical diagnosis aims for structured output to enhance the evaluation and readability of reports. ECG-Chat employs plasTeX \cite{plastex_documentation} to generate clinical diagnostic reports in a highly structured and automated manner. Initially, patient personal information, including basic details and key medical background, is collected and integrated from outpatient systems or patient databases, forming the "Patient Information" and "Medical History" sections. Subsequently, ECG-Chat analyzes ECG data to extract key metrics and patterns, which are detailed in the "ECG Data Analysis" section. The report then incorporates pathological analysis results, providing thorough descriptions and interpretations in the "Pathological Analysis" section. Based on the consolidated data and analyses, ECG-Chat generates the "Diagnosis" section, which explicitly states the clinical diagnosis. Finally, the system proposes specific treatment recommendations and preventive measures, which are outlined in the "Recommendations" section. A sample of this template is illustrated in Figure \ref{fig:report_1} and Figure \ref{fig:report_2}.

\section{Description of Datasets}
The following is a description of the open data set we used.
\begin{itemize}
    \item \textbf{MIMIC-IV-ECG:} This dataset contains 800,035 ECG records across nearly 160,000 unique patients \cite{mimic-ecg}. Every ECG recording sampled at 500Hz for a duration of 10 seconds and Each ECG corresponds to several machine-generated text reports. We process this dataset using the following strategy: (1) Replace 'NAN' and 'inf' with 0 in ECG data. (2) Text in the reports that are not relevant to the diagnosis are excluded, and samples whose final report is empty are removed. Finally, we use 788,822 samples as training data.
    \item \textbf{Champan-Shaoxing-Ningbo (CSN):} This dataset contains 45,152 12-lead ECG records \cite{csn_data1, csn_data2}. Every ECG recording is also sampled at 500Hz for a duration of 10 seconds. Each record is annotated by several SNOMED CT codes\footnote{ https://www.snomed.org/}. We converted each SNOMED CT code into a corresponding textual description, merged as a report of the ECG record, and used it for data training. The number of training samples is 40,637.
    \item \textbf{Shandong Provincial Hospital (SPH) database:} The dataset contains 25,770 ECG records from 24,666 patients \cite{sph_data}. The length of each recording is between 10 and 60 seconds, and the sampling rate is 500Hz. We intercept the first 10 seconds in the record as training data. Each recorded diagnosis in the dataset corresponds to standardized diagnostic statements conforming to the AHA/ACC/HRS recommendations \cite{aha_code}. Similarly, we also convert these statements into reports of corresponding records.  The number of training samples is 20,616.
    \item \textbf{PTB-XL: } This dataset contains 21,837 10s ECG signals that come from 18,885 patients. Each ECG record has a corresponding free text report and SCP Codes label \cite{scp_ecg} extracted from the report. 
    \item \textbf{CPSC2018: } This dataset has 6,877 standard 12-lead ECG records annotated by 9 distinct labels. The duration of the recording was 6-60s. We truncate those longer than 60s to 10s, and fill those shorter than 10s with zeros. The sampling rate in both dataset in 500Hz, which is consistent with the training data. The training and test sets of the two datasets are divided according to their respective official.
\end{itemize}

The following is how to create ECG-Chat pretraining and fine-tuning datasets.
\begin{itemize}
    \item \textbf{Pretraining dataset:} Due to the lack of large-scale annotated ECG datasets, we directly use the MIMIC-IV-ECG dataset as the training data for the feature alignment stage.Specifically, the pretraining dataset is derived from a subset of the MIMIC-IV-ECG dataset of size 619K. Each sample can be viewed as a single round conversation consisting of an ECG recording \(X_{e}\), a question \(X_{q}\) and an answer \(X_{a}\). \(X_{q}\) is a question about how to interpret the ECG, such as \textit{"Is there anything abnormal in this ECG?"}. To diversify the questions, we randomly generated 1.5K such questions using GPT-4o, randomly matched to the data in MIMIC-IV-ECG. \(X_{e}\) and \(X_{a}\) are the paired ECG-report in the original dataset. To make the responses more conversational, we integrated the reports of one ECG recording using the following format: \textit{"Your ECG shows \{Report 1\}; \{Report 2\}; ... . It's a normal/abnormal/borderline ECG."}
    \item \textbf{Fine-tuning dataset:} The dataset for this stage is ECG-Instruct, an instruction-following dataset built with language-only GPT-4o. The dataset contains two categories: diagnosis and conversations. The ECG records are both from the MIMIC-IV-ECG dataset. Fig. \ref{fig_ecg_ins} illustrates an example from the construction of the ECG-Instruct dataset.
    \begin{enumerate}
        \item \textit{Diagnosis:} Similar to the pretraining dataset, the diagnostic dataset is a single-round conversation dataset. The question is a request about interpreting an ECG and the answer is a detailed interpretation of a given ECG. To build the dataset, we gave GPT-4o the text reports of each ECG recording and a randomly selected question, making it summarize the reports and provide the answer. Compared to the pretraining data, the responses are more relevant to the user's instructions and like a doctor chatting to a patient. At the same time, this dataset ensures the accuracy of the answers. This dataset contains 19K ECG-question-answer samples. 
        \item \textit{Conversations:} This dataset was designed by GPT-4o for multi-turn ECG conversations between patients and physicians, including heart rate, waveform, rhythm, cardiac axis, and diagnostic results. Therefore, in addition to the text reports, the calculated waveform data by NeuroKit2 \cite{neurokit} on lead II were included in the prompt we provided to GPT-4o. Additionally, in order to enhance the robustness of the model, diagnostic or waveform features that did not exist in the original reports were randomly added to the prompt words, leading to negative answers in the data, as context type 3 shown in Fig. \ref{fig_ecg_ins}. This dataset contains 25K samples, each sample has 4-15 rounds dialogue. Of the 25K samples, 19K are from the diagnosis dataset.
    \end{enumerate}
\end{itemize}

We also used ECG-ExpertQA as the evaluation dataset for our knowledge base. This dataset contains 123 question-answer pairs automatically generated by ChatGPT-4o, guided by real medical cases and expert knowledge. Expert evaluation confirms that it effectively assesses the performance of large models in domain-specific knowledge.

\section{Dataset Statistics and Analysis on ECG-Instruct}

\begin{table}[h]
    \centering
    \setlength{\tabcolsep}{2pt}
    \begin{tabular}{c|c}
        \toprule
        \textbf{Indicators} & \textbf{Count} \\
        \midrule
        Number of vocabularies & 3,294,880 \\
        Number of distinctive vocabularies & 6,319 \\
        Number of sentences & 325,730 \\
        Average length of captions & 73.15 \\
        Average number of sentences per caption & 7.23 \\
        \midrule
        Number of ECGs & 45,044 \\
        \bottomrule
    \end{tabular}
    \caption{Statistical indicators of the ECG-Instruct dataset.}
    \label{tab:ecg_instruct_statistics_}
\end{table}

\begin{figure}[h]
        \centering
	\includegraphics[width=1\columnwidth]{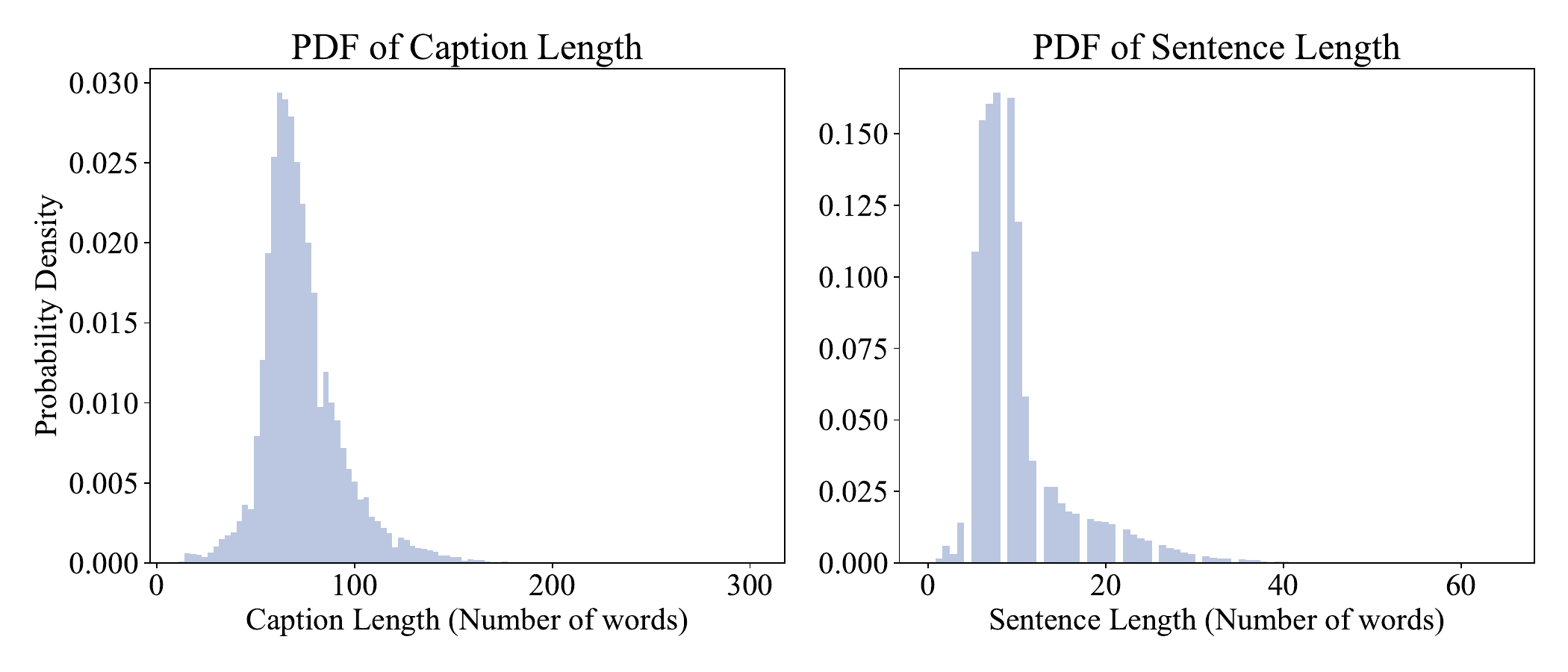} 
	\caption{The probability density function (PDF) visualization on ECG-Instruct}
	\label{ecg_instruct_statistics}
\end{figure}

Table \ref{tab:ecg_instruct_statistics_} and Figure \ref{ecg_instruct_statistics} present the statistical indicators of the ECG-Instruct dataset. The dataset contains 45,044 ECG samples with a total of 325,730 sentences, averaging 7.23 sentences per description. Each description has an average length of 73.15 vocabularies, amounting to a total of 3,294,880 vocabularies, including 6,319 distinct vocabularies. These statistics indicate that the ECG-Instruct dataset exhibits a high level of granularity in its textual descriptions, offering detailed explanations of ECG signals and their clinical implications. This rich textual information provides a solid foundation for understanding and automatically generating clinical diagnostic reports.
\begin{figure}[h]
        \centering
	\includegraphics[width=1\columnwidth]{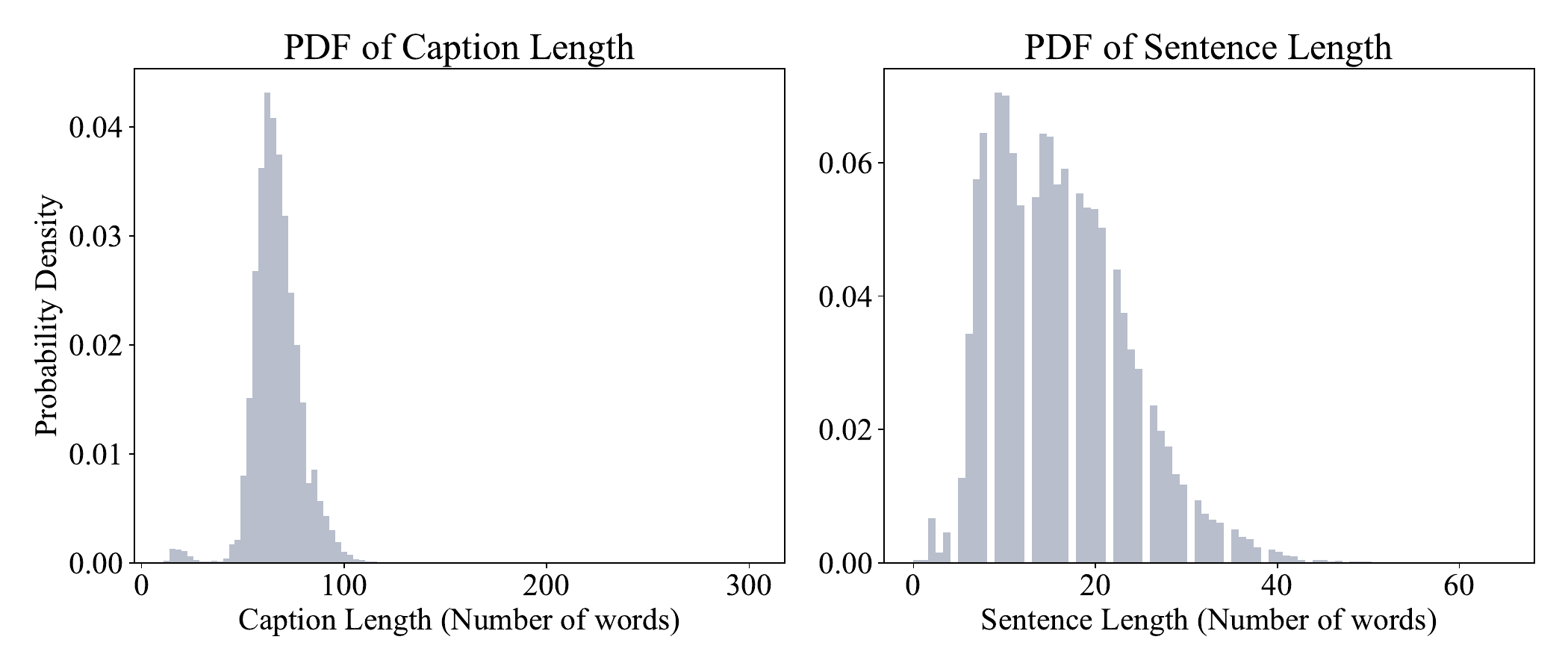} 
	\caption{The PDF visualization on ECG-Instruct-Diagnosis}
	\label{Diagnosis_statistics}
\end{figure}

Figure \ref{Diagnosis_statistics} displays the PDF for the Diagnosis component. This part of the dataset includes 20,258 ECG-question-answer samples, each representing a single-turn conversation where the question typically pertains to interpreting the ECG signal, and the answer provides a detailed explanation. The dataset features a total of 83,056 sentences with an average caption length of 66.77 words and an average of 4.10 sentences per caption. The vocabulary count includes 1,352,653 total words and 2,775 distinctive words. The range of caption lengths varies from a minimum of 11 words to a maximum of 303 words, and the number of sentences per caption ranges from 2 to 16. The figure illustrates the distribution of caption lengths, sentence counts, and vocabulary usage frequencies, reflecting the granularity and richness of this part of the dataset.
\begin{figure}[h]
        \centering
	\includegraphics[width=1\columnwidth]{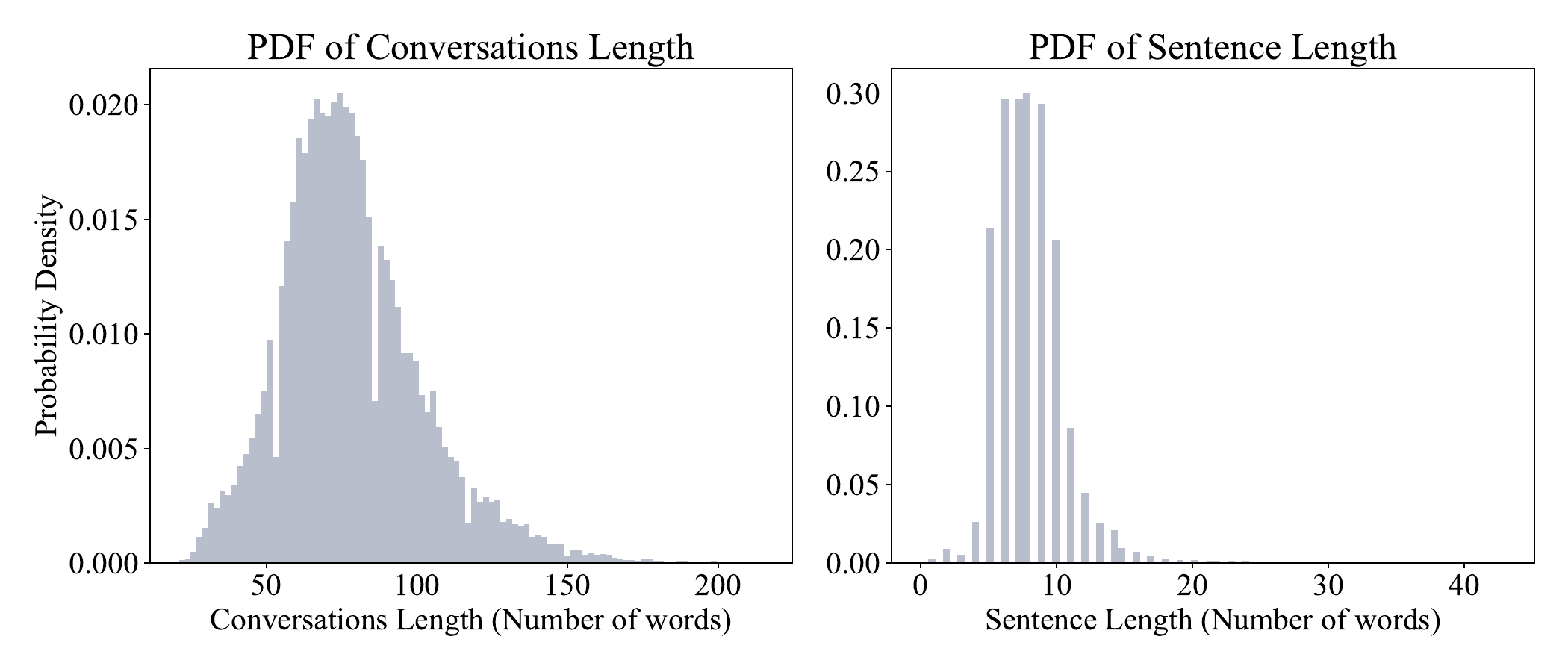} 
	\caption{The PDF visualization on ECG-Instruct-Conversations}
	\label{Conversations_statistics}
\end{figure}
\begin{figure*}[t]
        \centering
	\includegraphics[width=2.1\columnwidth]{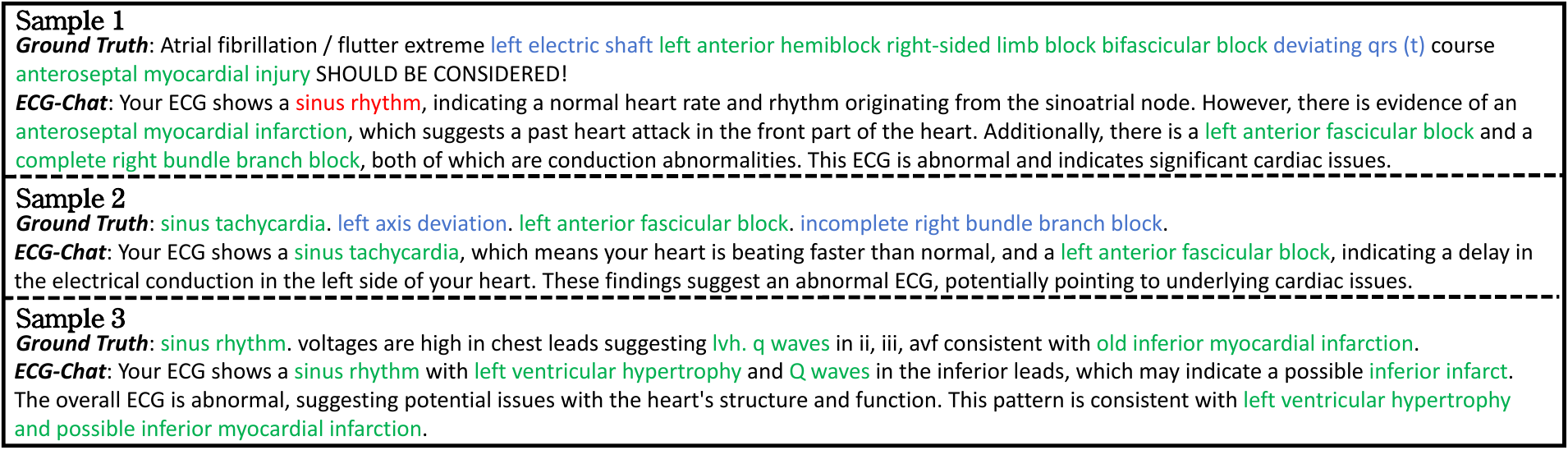} 
	\caption{Green highlights represent correct ECG findings, blue indicates findings omitted from the report, and red marks errors in the ECG-Chat interpretation compared to the Ground Truth. }
	\label{fig_report_sample}
\end{figure*}
Figure \ref{Conversations_statistics} displays the PDF for the Conversations component. This part contains 24,786 multi-turn dialogue samples, with each sample including 4 to 30 rounds of conversation covering topics such as heart rate, waveform, rhythm, cardiac axis, and diagnostic results. The dataset encompasses a total of 242,674 sentences, with an average conversation length of 78.36 words and an average of 9.79 sentences per conversation. The vocabulary count includes 1,942,227 total words and 4,774 distinctive words. The range of conversation lengths varies from a minimum of 21 words to a maximum of 215 words, and the number of sentences per conversation ranges from 4 to 30. The figure provides statistical information on conversation length distribution, sentence counts, and vocabulary usage, highlighting the diversity and complexity of this part of the dataset.

\section{Ablation Study}
\paragraph{Dataets}During the ECG encoder training process, we merged three training datasets. Table \ref{tab:ablation_dataset} compares the effects of a single MIMIC-IV-ECG dataset and a mixed dataset. It can be found that the mixed dataset can effectively enhance the generalization ability of the model.
\begin{table}[h]
    \centering
    \setlength{\tabcolsep}{2pt}
    \begin{tabular}{lcccc}
        \toprule
        & \multicolumn{2}{c}{Retrival (R@1)} & \multicolumn{2}{c}{Classification (F1)} \\
        \cmidrule(lr){2-3} \cmidrule(lr){4-5}
        Dataset & to report & to ECG & PTB-XL & CPSC2018 \\
        \midrule
        MIMIC-IV-ECG & 26.7 & 30.5 & 49.8 & 77.0 \\
        Mixed & \textbf{64.7} & \textbf{71.6} & \textbf{52.8} & \textbf{80.1} \\
        \bottomrule
    \end{tabular}
    \caption{Performance comparison between single dataset and mixed dataset.}
    \label{tab:ablation_dataset}
\end{table}

\begin{table}[h]
    \centering
    \setlength{\tabcolsep}{2pt}
    \begin{tabular}{lcccc}
        \toprule
        & \multicolumn{2}{c}{Retrival (R@1)} & \multicolumn{2}{c}{Classification (F1)} \\
        \cmidrule(lr){2-3} \cmidrule(lr){4-5}
        Model & to report & to ECG & PTB-XL & CPSC2018 \\
        \midrule
        BioLinkBert & \textbf{72.7} & \textbf{76.6} & 52.7 & 78.3 \\
        BioClinicalBert & 69.9 & 71.2 & 52.4 & 78.8 \\
        Med-CPT & 64.7 & 71.6 & \textbf{52.8} & \textbf{80.1} \\
        \bottomrule
    \end{tabular}
    \caption{Performance comparison between different text encoders.}
    \label{tab:ablation_text_encoder}
\end{table}

\paragraph{Text Encoder}We selected three different pretrained text encoders, Med-CPT \cite{medcpt} obtained on the contrastive learning task, and BioLinkBert \cite{linkbert} and BioClinicalBert \cite{clinicalbert} obtained on the reconstruction task. As shown in Table \ref{tab:ablation_text_encoder}, BioLinkBert and Med-CPT achieved the best results in retrieval and classification tasks respectively.

\begin{table}[h]
    \centering
    \caption{Scalability analysis results}
    \setlength{\tabcolsep}{2pt}

    \begin{subtable}[t]{\columnwidth}
        \centering

        \begin{tabular}{lcccc}
            \toprule
            & \multicolumn{2}{c}{Retrival (R@1)} & \multicolumn{2}{c}{Classification (F1)} \\
            \cmidrule(lr){2-3} \cmidrule(lr){4-5}
            Samples & to report & to ECG & PTB-XL & CPSC2018 \\
            \midrule
            80K & 0.86 & 1.09 & 45.42 & 66.6 \\
            402K & 40.1 & 49.4 & 51.2 & 75.6 \\
            805K & \textbf{64.7} & \textbf{71.6} & \textbf{52.8} & \textbf{80.1} \\
            \bottomrule
        \end{tabular}
        \caption{Performance comparison between different training samples.}
        \label{tab:ablation_train_num}
    \end{subtable}

    \vspace{1em} 

    \begin{subtable}[t]{\columnwidth}
        \centering

        \begin{tabular}{lcccc}
            \toprule
            & \multicolumn{2}{c}{Retrival (R@1)} & \multicolumn{2}{c}{Classification (F1)} \\
            \cmidrule(lr){2-3} \cmidrule(lr){4-5}
            Parameters & to report & to ECG & PTB-XL & CPSC2018 \\
            \midrule
            43M & 63.5 & 70.1 & \textbf{54.6} & 79.6 \\
            85M & 64.7 & 71.6 & 52.8 & 80.1 \\
            128M & \textbf{68.7} & \textbf{74.7} & 54.4 & \textbf{80.7} \\
            \bottomrule
        \end{tabular}
        \caption{Performance comparison between different number of parameters of ECG encoders.}
        \label{tab:ablation_model_parameter}
    \end{subtable}
\end{table}

\paragraph{Scalability}We train the model on 10\%, 50\% and 100\% of the data, as shown in Table \ref{tab:ablation_train_num}. Using all the data, we train models with 43M, 85M and 128M parameters, respectively, by changing the number of transformer layers. The results are shown in Table \ref{tab:ablation_model_parameter}. It can be seen that the number of model parameters has little impact on the results. When increasing the amount of training data, the model effect can be significantly improved, especially on retrieval tasks.

\begin{table*}[h]
\centering
\setlength{\tabcolsep}{10pt}
\renewcommand{\arraystretch}{1.2}
\begin{tabular}{cc|ccccccc}
\toprule
GraphRAG & DSPy & F & AR & CR & CP & CU & CER & SS \\
\midrule
\centering \ding{56} & \centering \ding{56} & 39.87 & 32.56 & 9.03 & 29.82 & 29.13 & 24.67 & 18.94 \\
\centering \ding{52} & \centering \ding{56} & 76.60 & 68.29 & 32.67 & 67.53 & 64.39 & 57.00 & 54.40 \\
\centering \ding{56} & \centering \ding{52} & 71.57 & 63.93 & 27.57 & 62.81 & 59.54 & 53.35 & 47.72 \\
\centering \ding{52} & \centering \ding{52} & \textbf{82.12} & \textbf{74.29} & \textbf{39.44} & \textbf{73.18} & \textbf{88.92} & \textbf{73.10} & \textbf{81.83} \\
\bottomrule
\end{tabular}
\caption{Ablation Study of GraphRAG and DSPy in Model Performance Metrics}
\label{tab:GraphRAG_and_DSPy}
\end{table*}
\paragraph{Effectiveness of GraphRAG and DSPy}
We designed an ablation experiment to evaluate the practical effects of GraphRAG and DSPy in the ECG-Chat model. Using a small-scale question-answer dataset containing 123 complex questions, ECG-ExpertQA, we compared the model's performance with and without these two modules. To this end, we employed RAGAS as the evaluation tool, focusing on the model's performance across key metrics including Faithfulness (F), Answer Relevancy (AR), Context Recall (CR), Context Precision (CP), Context Utilization (CU), Context Entity Recall (CER), and Summarization Score (SS).

Table \ref{tab:GraphRAG_and_DSPy} shows the combination of GraphRAG and DSPy significantly outperforms other combinations across all metrics. Without GraphRAG and DSPy, the model's performance is relatively poor, with a Faithfulness score of 39.87 and Answer Relevancy score of 32.56, among others. Introducing the GraphRAG module results in a significant improvement in Faithfulness (76.60) and Answer Relevancy (68.29). Meanwhile, incorporating the DSPy module leads to improvements in Context Recall (27.57) and Summarization Score (47.72).
\section{Case Study}
Figure \ref{fig_report_sample} shows some reports generated by ECG-Chat and compared with the translated reports in PTB-XL. It can be seen that the report text of ECG-Chat is more fluent and colloquial, suitable for patients to read. At the same time, it has a relatively good diagnostic accuracy.

\end{document}